\RequirePackage[hyphens]{url} 
\documentclass[twocolumn]{aastex63}

\renewcommand{\today}{2020 October 21} 

\usepackage{microtype}
\usepackage{amsmath}
\usepackage{mathtools} 

\let\tablenum\relax 
\usepackage[separate-uncertainty]{siunitx}

\providecommand{\sorthelp}[1]{} 

\usepackage{graphicx}

\usepackage{xcolor}
\definecolor{plotblue}{RGB}{0, 107, 164}
\definecolor{plotgray}{gray}{0.54}
\definecolor{plotgreen}{RGB}{100, 200, 100}

\usepackage{bbding} 

\usepackage{soul}
\usepackage[outline]{contour}
\newcommand \colorindicator[3]{%
  \begingroup%
  \setul{0.28ex}{0.4ex}%
  \contourlength{0.2ex}%
  \setulcolor{#1}%
  \ul{{\phantom{#2}}}\llap{\contour{white}{#2}}\if\relax\detokenize{#3}\relax\else~\textcolor{#1}{\tiny{#3}}\fi%
  \endgroup%
}

\begin{document}

\title{Full-sky Cosmic Microwave Background Foreground Cleaning Using Machine Learning}
\author[0000-0002-4436-4215]{Matthew~A. Petroff}
\author[0000-0002-2147-2248]{Graeme~E. Addison}
\author[0000-0001-8839-7206]{Charles~L. Bennett}
\author[0000-0003-3017-3474]{Janet~L. Weiland}
\affiliation{Department of Physics \& Astronomy, Johns Hopkins University, Baltimore, MD 21218, USA}

\shorttitle{CMB Foreground Cleaning Using Machine Learning}
\shortauthors{Petroff et al.}

\correspondingauthor{Matthew A. Petroff}
\email{petroff@jhu.edu}

\keywords{\href{http://astrothesaurus.org/uat/322}{Cosmic microwave background radiation (322)}; \href{http://astrothesaurus.org/uat/1146}{Observational cosmology (1146)}; \href{http://astrothesaurus.org/uat/1938}{Convolutional neural networks (1938)}}

\begin{abstract}
\noindent In order to extract cosmological information from observations of the millimeter and submillimeter sky, foreground components must first be removed to produce an estimate of the cosmic microwave background (CMB). We developed a machine-learning approach for doing so for full-sky temperature maps of the millimeter and submillimeter sky. We constructed a Bayesian spherical convolutional neural network architecture to produce a model that captures both spectral and morphological aspects of the foregrounds. Additionally, the model outputs a per-pixel error estimate that incorporates both statistical and model uncertainties. The model was then trained using simulations that incorporated knowledge of these foreground components that was available at the time of the launch of the Planck satellite. On simulated maps, the CMB is recovered with a mean absolute difference of \SI{<4}{\micro\kelvin} over the full sky after masking map pixels with a predicted standard error of \SI{>50}{\micro\kelvin}; the angular power spectrum is also accurately recovered. Once validated with the simulations, this model was applied to Planck temperature observations from its \SI{70}{\giga\hertz} through \SI{857}{\giga\hertz} channels to produce a foreground-cleaned CMB map at a \textsc{Healpix} map resolution of \textsc{nside}=512. Furthermore, we demonstrate the utility of the technique for evaluating how well different simulations match observations, particularly in regard to the modeling of thermal dust.
\end{abstract}

\section{Introduction}

The cosmic microwave background (CMB), a relic of the early Universe, encodes information about the physics of the first moments of the Universe. This wealth of information has proven crucial to modern precision cosmology. However, we can only observe the CMB through the local contamination of the Milky Way and other galaxies in the intervening Universe, which has fluctuations that are many times brighter than the anisotropy of the primordial signal \citep{Planck2018}. Fortunately, the spectral indices of the foregrounds, both Galactic and extragalactic, differ from that of the CMB, allowing for the signals to be discriminated between through observations made at multiple frequencies. The most straightforward method to do so is to perform a simple internal linear combination of sky maps observed at different frequencies, fitting templates to the various components based on their spectral indices \citep{Bennett2003}. Numerous other more advanced foreground cleaning techniques have also been developed in an attempt to more completely remove foregrounds and reduce systematic errors \citep[e.g.,][]{Martinez-Gonzalez2003, Hansen2006, Bobin2007, Cardoso2008, Eriksen2008, Katayama2011, Remazeilles2011, Basak2012, Ichiki2019}. Reduction of systematic errors is crucial as such errors bias the cosmological parameters extracted from the CMB map.

Machine learning, specifically deep learning using artificial neural networks, has been increasingly applied to astrophysical problems \citep[see reviews in][]{Ball2010, Carleo2019, Fluke2019}, including to the CMB \citep[e.g.,][]{Auld2008, Ciuca2017, He2018, Caldeira2019, Munchmeyer2019, Puglisi2020, VafaeiSadr2020, Yi2020}. Of particular note are previous applications of neural networks to CMB foreground cleaning, but these approaches have either not taken into account morphological features \citep{Norgaard-Nielsen2010} or have not been applied to the full sky \citep{Aylor2019}. An artificial neural network is a neuroscience-inspired construct that consists of a set of units connected with trainable weights and nonlinear transformations, which can be used to approximate arbitrary functions \citep{Russell2010}. Convolutional neural networks (CNNs) are a type of artificial neural network particularly well suited for cleaning foregrounds from maps, as they work on images and process structural information instead of only individual pixels \citep{Krizhevsky2012}. However, traditional CNNs are limited to rectangular images and thus are not well suited to full-sky maps of the CMB, although they can be applied to flat sky projections of limited parts of the sky. Recently, multiple techniques have been developed to extend CNNs to the sphere, allowing for the full sky to be considered at once \citep{Cohen2018, Esteves2018, Kondor2018, Jiang2019, Krachmalnicoff2019, Perraudin2019}. We apply one such technique to cleaning foregrounds from full-sky temperature maps of the millimeter and submillimeter sky, leaving the primordial CMB signal, as well as possible residuals. We also assess its ability to evaluate how well different simulations match observations.

The remainder of this paper is organized as follows. First, in Section~\ref{sec:architecture}, we discuss the artificial neural network architecture used. Next, in Section~\ref{sec:data}, we discuss the simulations used to train the model. We then describe the training procedure in Section~\ref{sec:training} and apply the trained model to the simulations. The trained model is then applied to Planck sky maps in Section~\ref{sec:planck}. Finally, we conclude in Section~\ref{sec:conclusion}.

\section{Neural network architecture}
\label{sec:architecture}

In concordance with the cosmological principle, a neural network architecture for cosmological applications should ideally encode neither position nor direction and should encompass the entire sky. The \emph{DeepSphere} architecture of \citet{Perraudin2019} fulfills these properties; it is a convolutional architecture, so it does not encode position, and it learns radially symmetric kernels, so it is approximately rotationally equivariant,\footnote{The improved weighting scheme presented in \citet{Defferrard2020} was used.} all while covering the full sky. This architecture represents the \textsc{Healpix} equal-area sphere pixelization \citep{Gorski2005} as a graph. Convolutions are performed as operations on the graph using Chebyshev polynomials, while downscaling takes advantage of \textsc{Healpix}'s hierarchical nature, combining four pixels into their single parent pixel. However, this is not a complete solution, since the technique, as presented by \citet{Perraudin2019}, only produces outputs suitable for regression or classification problems.

CMB foreground cleaning falls under the category of image-to-image problems in the machine-learning parlance. Thus, we extend the widely used \emph{U-Net} architecture of \citet{Ronneberger2015} to the sphere. This technique consists of both contracting and expanding paths, which decrease and increase the image resolution, respectively, in order to encode features on multiple scales; \emph{skip connections} are used to bypass smaller levels of the contracting--expanding path to transfer details and bypass the information bottleneck of the smallest level. On the contracting path, after a convolution is applied, a pooling operation is used to combine a block of four pixels into a single pixel; in this case, we use \emph{max pooling}, which takes the maximum value of the four input pixels. On the expanding path, each pixel is subdivided into four identical child pixels, and a convolution is then performed on the higher resolution image. Each convolution layer uses a set of trainable filters and is generally followed by a nonlinear activation layer.

To extend this architecture to the \textsc{Healpix} sphere, we use \textsc{Healpix}'s hierarchical nature to either combine four pixels into a single pixel or to subdivide a pixel into four child pixels, depending on whether the resolution is being decreased or increased. For the contracting path, we apply a DeepSphere graph-based convolution before pooling, while for the expanding path, we apply a DeepSphere graph-based convolution after pixels are duplicated to produce a higher resolution map.

\begin{figure*}
\centering
\includegraphics[width=\textwidth]{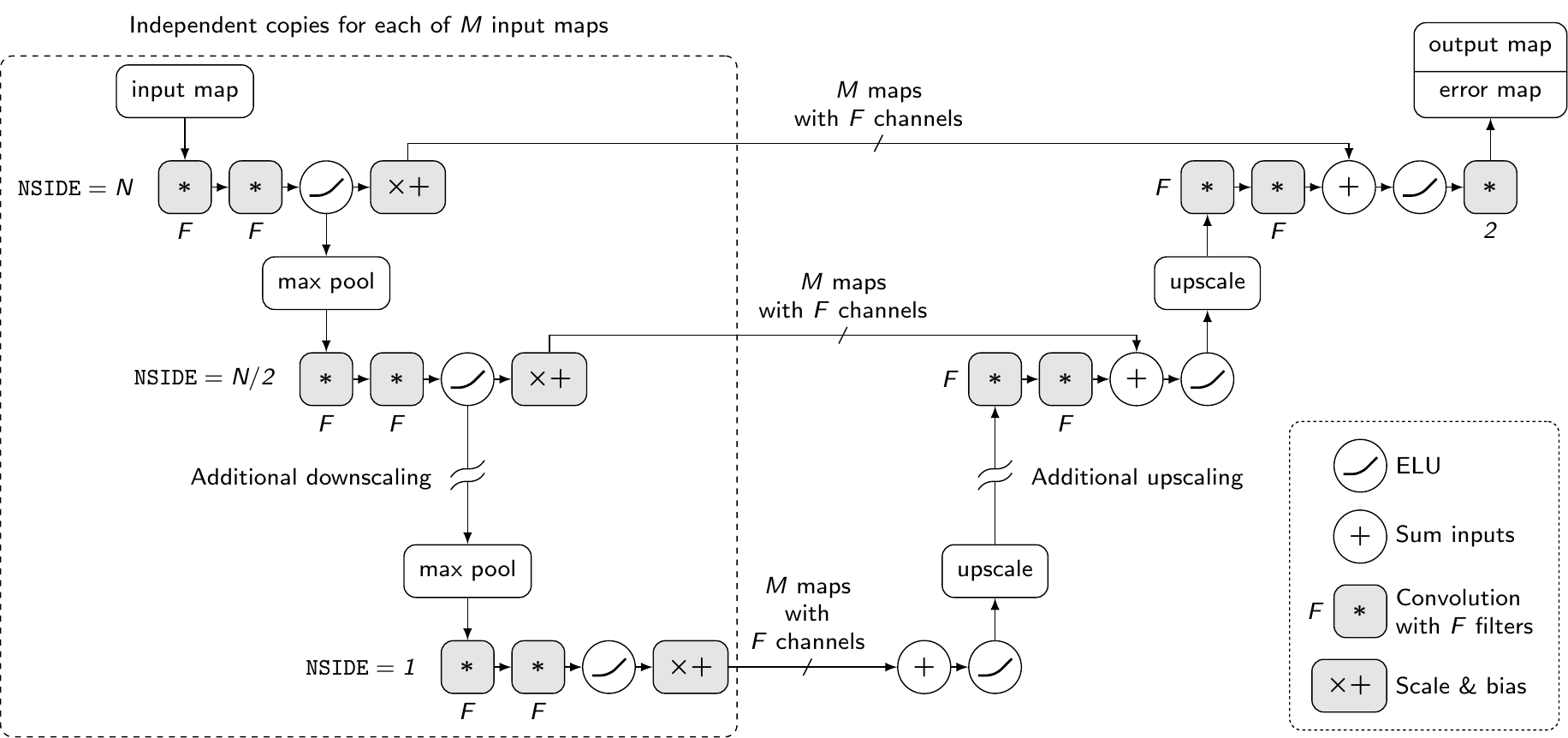}
\caption{Artificial neural network architecture overview. Shading denotes nodes with trainable parameters.}
\label{fig:ann}
\end{figure*}

Finally, a method of combining features from separate frequency maps is needed. We accomplish this by using a separate contracting path for each of the input frequency maps, with trainable multiplicative scale and additive bias weights for each frequency at each detail level. While a single contracting path that takes all of the frequency channels as input could have been used with a greater number of convolution filters, this would have resulted in a larger model with a greater number of trainable parameters. Instead, we exploit a priori knowledge of astrophysical foregrounds and construct an architecture that allows the model to learn resolution-dependent internal linear combinations in addition to morphological features. On the single expanding path, the scaled and biased frequency maps are summed with the upscaled, convolved map from the smaller level in the path. After this step, a nonlinear activation is applied to the summed map, and it is upscaled and convolved. The process is repeated for the more detailed result. A schematic overview of this architecture is shown in Figure~\ref{fig:ann}. All nonlinear activations are performed using the exponential linear unit (ELU) operation \citep{Clevert2016}. The number and sizes of the convolution filters will be discussed in Section~\ref{sec:training}. Each location where a convolution is performed consists of a pair of convolution layers, since this was found to reduce the loss function (defined below) more than a single convolution layer with twice the number of filters (but the same total number of filters).

For machine learning to be maximally useful in scientific contexts, accurate error estimation is crucial. This estimated error can be divided into aleatoric uncertainty, uncertainty due to noise in the observations, i.e., statistical uncertainty, and epistemic uncertainty, uncertainty due to error in the model, i.e., model-based systematic uncertainty \citep{Kendall2017}. To this end, we used a Bayesian neural network with \emph{Concrete dropout} \citep{Gal2017}. \emph{Dropout} is a process that randomly drops units, and their connections, from an artificial neural network during training, which was developed to prevent overfitting during the training process \citep{Srivastava2014}. Dropout was later shown to be a Bayesian approximation that can be used to estimate a network's epistemic uncertainty by also randomly dropping units over multiple evaluations, but the dropout probability had to be tuned to calibrate the uncertainty estimate \citep{Gal2016}; Concrete dropout allows for automatic tuning of the dropout probability. Following \citet{Kendall2017}, we have the network estimate aleatoric uncertainty by producing an error map in addition to a CMB map. This is accomplished by training the network to minimize the loss function
\begin{equation}
\label{eq:loss}
\mathcal{L} = \frac{1}{D}\sum_{i=0}^D\left[\frac{1}{2}\exp(-\log\hat{\sigma}_i^2)||\mathbf{y}_i - \hat{\mathbf{y}}_i||^2+\frac{1}{2}\log\hat{\sigma}_i^2\right],
\end{equation}
where $\mathbf{y}$ is the true (simulated) CMB map, $\hat{\mathbf{y}}$ is the predicted CMB map, and $\hat{\sigma}^2$ is the predicted variance, all averaged over $D$ pixels. As per \citet{Kendall2017}, the log variance is used as the model output for numerical stability reasons. The two types of uncertainty are combined by evaluating the network many times with dropout. Each iteration, a random noise realization is drawn from a Gaussian distribution with variance corresponding to the per-pixel aleatoric uncertainty estimates, which is added to the CMB map realization. The mean and variance of this set of map realizations are then calculated to produce the final output.

\section{Training data}
\label{sec:data}

In order to train a neural network model, many training samples are required. Since we have only one sky to observe, and since we do not know the ground truth for that sky, simulations must be used. For this, the pre-launch \emph{Planck Sky Model} was used \citep{Delabrouille2013}. This mostly template-based model includes diffuse synchrotron, thermal dust, free--free, spinning dust, and CO foreground components, as well as the Sunyaev--Zel$^\prime$\!dovich effect for clusters and the cosmic infrared background (CIB), using the knowledge of these components that was available when the Planck spacecraft was launched. Although the Planck Sky Model can also simulate strong and weak point sources, these were excluded from the simulations that were run, since we found the inclusion of point sources to cause issues with the neural network training; a catalog of point sources can be used to mask the final output map before angular power spectrum estimation instead, if desired, but we did not find this masking to be necessary. The Planck Sky Model generates randomized constrained realizations of these foreground components, which match prior observations to within observational errors. As the uncertainty on the knowledge of these components is included in the simulations, this can be used to help better calibrate the uncertainties in the Bayesian neural network model. Furthermore, it allows for creating an independent foreground cleaning technique, since no Planck data are used in the simulations, except for derived instrument parameters. For reasons that will be discussed in Section \ref{sec:dustmodels}, the thermal dust simulation in the Planck Sky Model code was modified to use Model 8 of \citet{Finkbeiner1999} instead of the default of Model 7.\footnote{Contrary to the description in \citet{Delabrouille2013}, the pre-launch Planck Sky Model (v1.7.8) does not include Model 8 as an option.}

To generate the CMB used with the simulations, the CLASS Boltzman code \citep{Blas2011} was used with parameters drawn from the WMAP9 flat $\Lambda$CDM Monte Carlo chains \citep{Hinshaw2013} to produce theory angular power spectra, some of which deviate significantly from current constraints; thus, the simulated spectra cover the range of plausible spectra. These spectra were then modified to account for CMB lensing, and the lensed spectra were used to generate sky map realizations. The simulated sky was then observed using the Planck 2018 reduced instrument model \citep{Planck2018}, without including instrument noise or instrument beams, using temperature data from both the \SI{70}{\giga\hertz} low frequency instrument (LFI) band and all six high frequency instrument (HFI) bands. A resolution of \textsc{nside}=512 was used,\footnote{Early experimentation was performed with maps with resolution \textsc{nside}=64 or \textsc{nside}=256, since this was considerably faster.} and the sky was smoothed to a common \SI{13.1}{\arcminute} Gaussian beam, corresponding to the largest Planck beam used, that of the \SI{70}{\giga\hertz} channel. Finally, the $TT$ variance maps included with the published Planck 2018 frequency maps were used to add white instrument noise to the simulated frequency maps. In addition to the observed frequency maps, the reference CMB map was also saved, to be used as the reference during model training. A set of 1000 simulations was created. Using the full simulation set, a set of normalization values were derived and applied multiplicatively to keep each frequency map between $-1$ and $1$ and each CMB map between $0$ and $1$, with the CMB maps centered at 0.5.

\section{Training procedure}
\label{sec:training}

For training, the set of 1000 simulations was subdivided into a training set consisting of 800 simulations and a test set containing the remaining 200 simulations. As mentioned in the previous section, these simulations differ in their randomized constrained realizations of foreground components, CMB components (including $\Lambda$CDM parameters), and instrument noise realizations. The network architecture was implemented using \emph{TensorFlow} \citep{Abadi2016}, and the training procedure was performed on an NVIDIA TITAN RTX GPU. For training, the \emph{AMSGrad} optimizer \citep{Reddi2018} was used with default parameters. Six filters were used for each convolution layer (parameter $F$ in Figure~\ref{fig:ann}), with a polynomial order of nine; increasing either parameter was found to only provide small improvements, while thus increasing the network complexity unnecessarily. For Concrete dropout, a length scale of $10^{-4}$ was used. The final model includes \num{50729} trainable parameters.

The model was trained for 400 epochs,\footnote{An epoch refers to a single pass through the training dataset during the training procedure.} with a batch size of one,\footnote{The batch size denotes how many examples are used simultaneously during the training procedure.} a process that took approximately nine days. The \emph{TensorFlow Large Model Support} library \citep{Le2019}\footnote{\url{https://github.com/IBM/tensorflow-large-model-support}} was used to allow portions of the model to be swapped out of GPU memory during the training process, since the model could not otherwise be trained, even with a batch size of one, due to GPU memory limitations. After training, the average loss [equation~(\ref{eq:loss})] on the training simulations was $-7.22$, while the average loss on the test simulations was also $-7.22$. This shows that the model was not overfit to the training examples, although this does not eliminate the possibility of fitting to shortcomings in the simulations relative to reality.

\begin{figure*}
\centering
\includegraphics[width=0.87\textwidth]{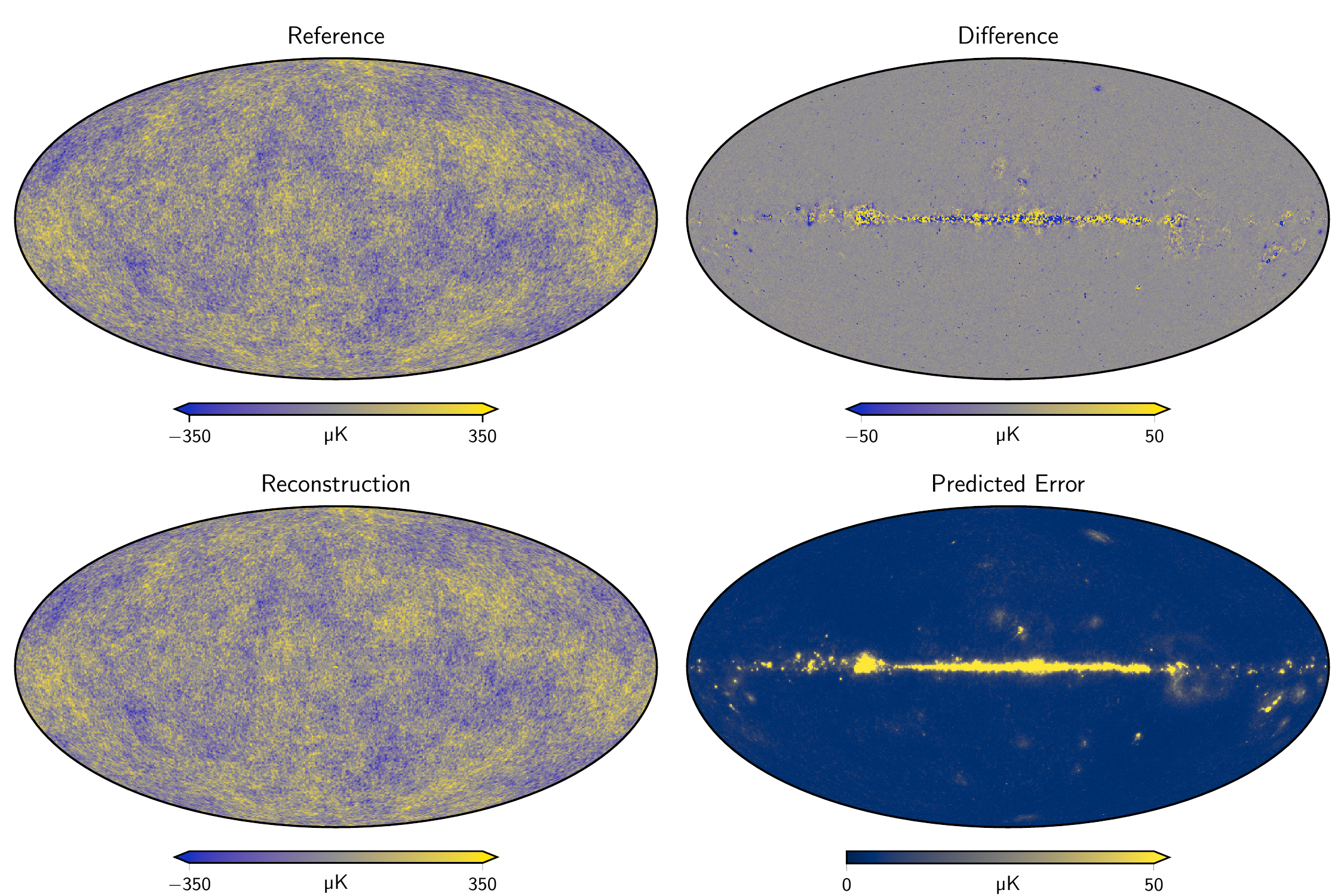}
\caption{Simulated maps from one example in the test set. The top left panel shows the reference CMB map, the bottom left panel shows the CMB as reconstructed by the neural network model, the top right panel shows the difference between the reference and reconstructed CMB maps, and the bottom right panel shows the error predicted by the neural network model. Note that the color scales for the difference and predicted error saturate on portions of the Galactic plane.}
\label{fig:testmaps}
\end{figure*}

The results of evaluating the trained model for one set of maps from the test set are shown in Figure~\ref{fig:testmaps}, utilizing the average of 100 evaluations to predict the uncertainty using the previously mentioned procedure of \citet{Kendall2017}. The reconstructed map shows only small residuals away from the Galactic plane (\SI{<4}{\micro\kelvin}, on average), and the larger residual directly on the plane ($\sim$\SI{40}{\micro\kelvin}, on average) is accurately captured by the predicted error. For this particular example, the mean predicted standard error and mean absolute difference over the full sky are \SI{5.4}{\micro\kelvin} and \SI{3.8}{\micro\kelvin}, respectively, after masking map pixels with a predicted standard error \SI{>50}{\micro\kelvin}; for 75\% of pixels, the actual difference was within the predicted standard error. These statistics show that the predicted error is close, although slightly overestimated.

\begin{figure}
\centering
\includegraphics[width=\columnwidth]{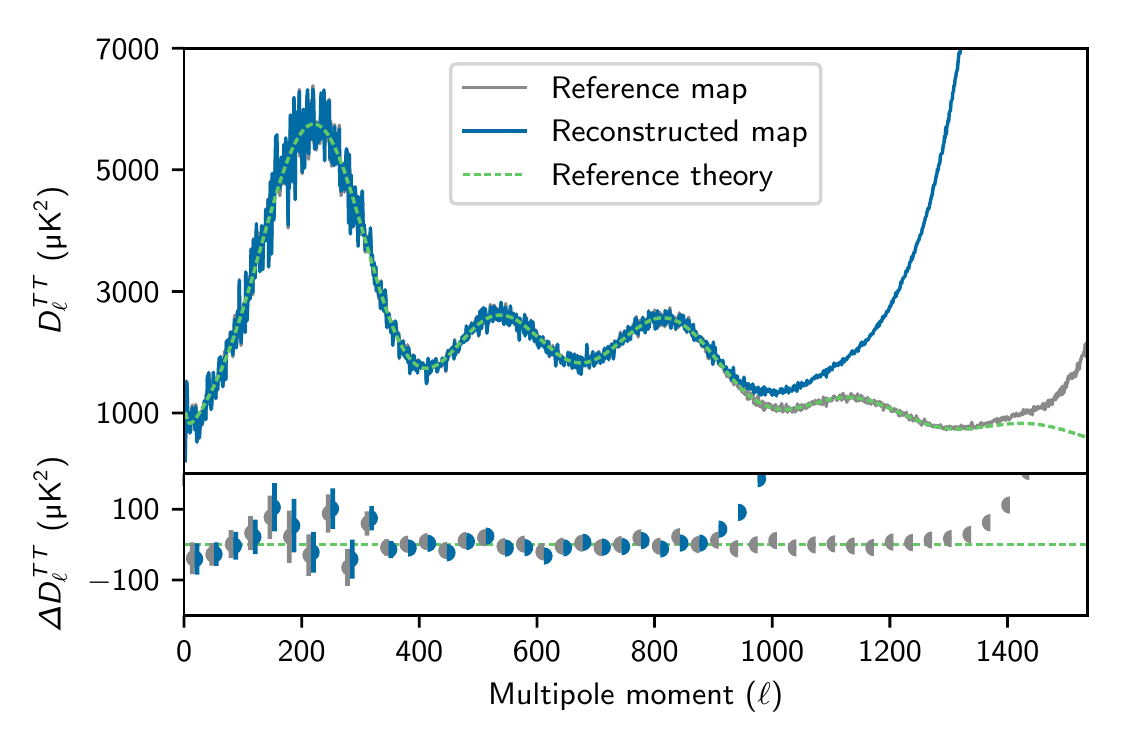}
\caption{Temperature angular power spectra of one example of simulated maps from the test set. In the top panel, the spectrum calculated from the reference map is shown in \colorindicator{plotgray}{gray}{\HalfCircleLeft}, the spectrum calculated from the reconstructed map is shown in \colorindicator{plotblue}{blue}{\HalfCircleRight}, and the theory spectrum used to create the simulations is shown in \colorindicator{plotgreen}{green}{}. The reference map spectrum is calculated without a mask or weighting, while the reconstructed map spectrum is calculated using the mask and weighting scheme described in the text. The bottom panel shows the mean and standard deviation of the difference between the map-derived spectra and the theory spectrum, using a bin width of 33, for both maps.}
\label{fig:testspectra}
\end{figure}

The temperature angular power spectrum of the reconstructed map was also evaluated using the \emph{PolSpice} estimator \citep{Chon2004}, with pixels with a predicted standard error \SI{>50}{\micro\kelvin} masked out and inverse variance weighting used for the remaining pixels, using the inverse of their predicted errors. While the reconstructed map shown in Figure~\ref{fig:testmaps} is the pixel-space average of the 100 individual maps output by the model, including noise drawn from their predicted errors, the angular power auto-spectrum was instead calculated for each individual map after the predicted noise was added;\footnote{A common mask and weighting were used for all the maps.} the resulting spectra were then averaged. This procedure was used as it produced improved results at smaller angular scales, when compared to the auto-spectrum calculated from the averaged map or to the average of cross-spectra calculated between the individual maps output by the model. The recovered spectrum matches up until $\ell\approx 900$, after which beam and noise effects start causing the spectra to diverge; the Planck \SI{70}{\giga\hertz} channel becomes noise-dominated for $\ell\gtrsim 800$ \citep{Planck2018lfi}, so this is a likely cause of the lack of signal at these smaller angular scales. Beam effects also affect the reference map, since it is also smoothed with the common \SI{13.1}{\arcminute} beam, but to a lesser degree. The above procedure was also applied to 100 map sets from the training set, and the resulting spectra were compared to spectra computed from the reference maps, to check for bias. A small multiplicative bias ($\lesssim 3\%$) was found; this was corrected by using the results of a quadratic fit to the bias for $2 < \ell < 900$, which resulted in a mean absolute residual bias of $\sim 0.3\%$ for $2 < \ell < 900$. The need for this correction is thought to be primarily due to the use of a pixel-space loss function during model training; however, it also effectively removes any noise bias in the $\ell$ range it is fit to. The resulting angular power spectrum of the test set example is shown in Figure~\ref{fig:testspectra}, along with the spectrum of the reference map and the theory spectrum used to generate the reference map and the simulated frequency maps.

\section{Application to Planck observations}
\label{sec:planck}

With the neural network model trained, it was then applied to the Planck 2018 intensity frequency maps, which were first corrected to use the same CIB monopole value as used in the simulations and normalized with the same multiplicative constants used for the simulations. The same evaluation procedure previously described for the map set from the simulation test set was used, with one caveat. The resulting reconstructed map has a slight monopole bias of \SI{\sim 1}{\micro\kelvin}, which is corrected for by masking \SI{\pm 10}{\degree} from the Galactic plane, calculating the median, and subtracting the median value from the map. This correction was implemented during development of the technique, when the bias was noticeably larger; while no longer necessary, the correction was kept for the sake of robustness. The bias is thought to result from the model using slightly different weights for each of the Planck frequency maps than when the simulated maps are used. When the frequency maps do not share a common beam window function, the power spectrum is also biased for the Planck data, unlike for the simulations, which is thought to be due to the same effect. This led to the decision to use a common beam window function for the frequency maps; the \SI{30}{\giga\hertz} and \SI{44}{\giga\hertz} LFI maps were thus excluded, due to their larger beam sizes. Although additionally excluding the \SI{70}{\giga\hertz} LFI map would allow for use of an even smaller beam, this benefit was found to be outweighed by the reduced ability to exclude synchrotron foregrounds.

\begin{figure*}
\centering
\includegraphics[width=0.87\textwidth]{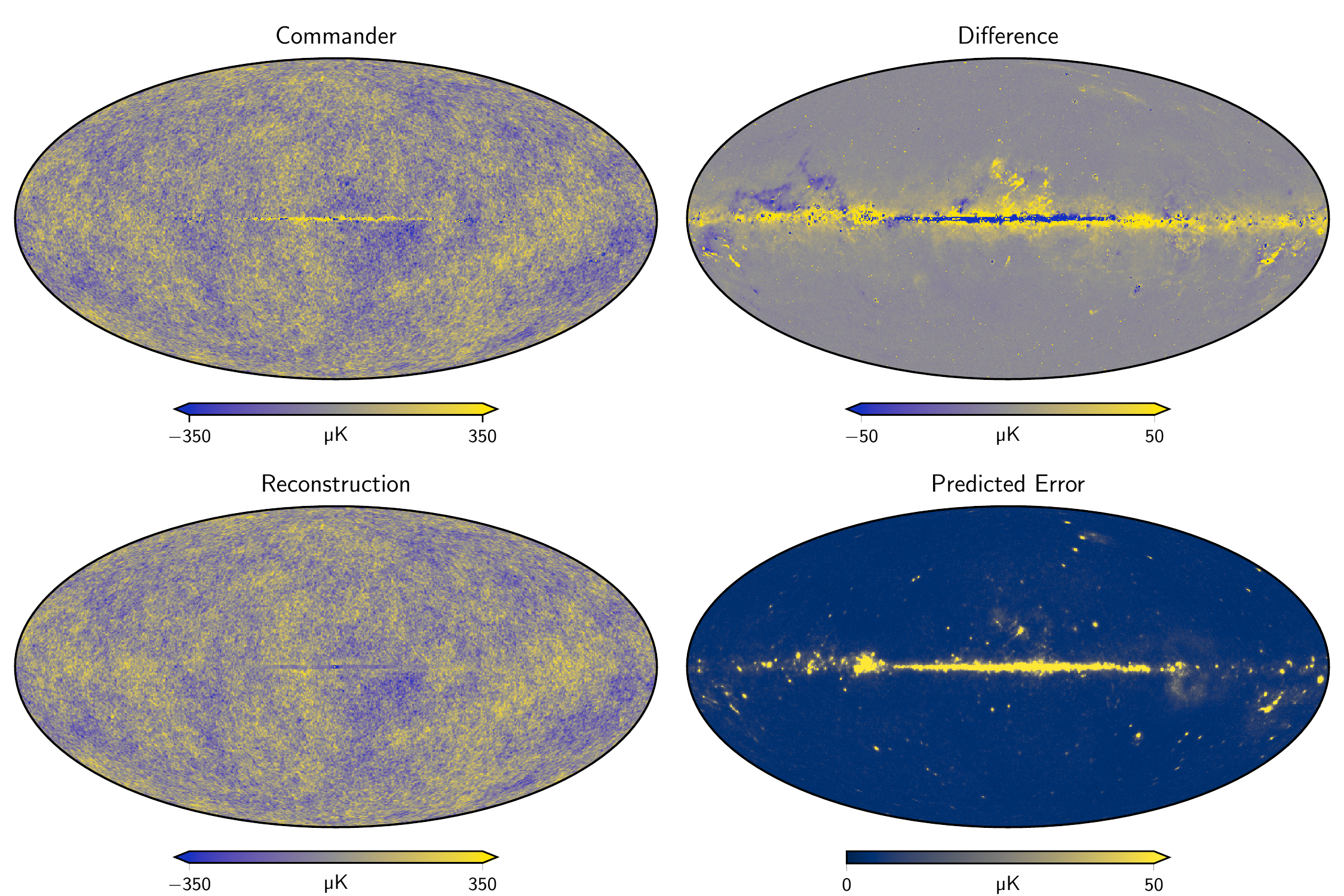}
\caption{Planck maps. The top left panel shows the Planck Commander CMB map, the bottom left panel shows the CMB as reconstructed by the neural network model from the Planck frequency maps, the top right panel shows the difference between the Commander map and the neural network reconstructed CMB map, and the bottom right panel shows the error predicted by the neural network model. Note that the color scales for the difference and predicted error saturate on portions of the Galactic plane.}
\label{fig:planckmaps}
\end{figure*}

The resulting reconstructed CMB map is shown in Figure~\ref{fig:planckmaps}, with comparison to the Planck Commander foreground-cleaned map \citep{Planck2018sep}, smoothed with the same beam window function. A residual structure like that of a diffuse Galactic foreground component is apparent, although it cannot be conclusively attributed to a specific foreground cleaning technique, since the ground truth is unknown. The mean predicted standard error is \SI{5.5}{\micro\kelvin}, after masking map pixels with a predicted standard error \SI{>50}{\micro\kelvin}. With the same masking, the mean absolute difference between the reconstructed map and the Commander map is \SI{6.2}{\micro\kelvin}, which is similar to the differences between the various Planck foreground cleaning techniques; the Commander--NILC, Commander--SEVEM, and Commander--SMICA mean absolute differences are \SI{5.5}{\micro\kelvin}, \SI{6.6}{\micro\kelvin}, and \SI{4.1}{\micro\kelvin}, respectively, when smoothed with the previously used beam window function. Since the predicted standard error matches the mean absolute difference with the Commander map, the error may again be overestimated, as the Commander map also has errors. However, since some of the errors in the two maps might be correlated, a definitive conclusion on the error estimation accuracy cannot be reached.

\begin{figure}
\centering
\includegraphics[width=\columnwidth]{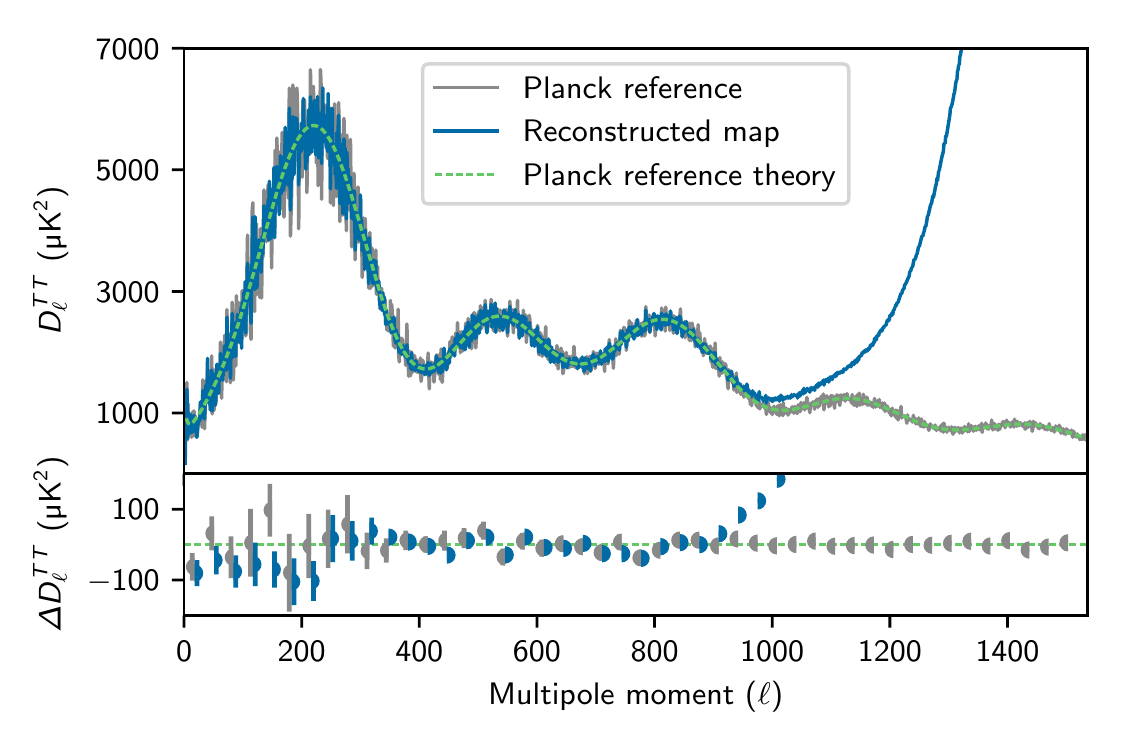}
\caption{Planck temperature angular power spectra. In the top panel, the published Planck spectrum is shown in \colorindicator{plotgray}{gray}{\HalfCircleLeft}, the spectrum calculated from the neural network reconstructed map is shown in \colorindicator{plotblue}{blue}{\HalfCircleRight}, and the published Planck best-fit theory spectrum is shown in \colorindicator{plotgreen}{green}{}. The reconstructed map spectrum is calculated using the mask and weighting scheme described in the text. The bottom panel shows the mean and standard deviation of the difference between the map-derived spectrum and the published Planck best-fit theory spectrum, as well as the difference between the published Planck spectrum and the  published Planck best-fit theory spectrum, using a bin width of 33.}
\label{fig:planckspectra}
\end{figure}

The temperature angular power spectrum of the reconstructed Planck map was also evaluated using the same technique that was used for the map set from the simulation test set, results of which are shown in Figure~\ref{fig:planckspectra}, with comparison to the published Planck spectrum and the published Planck best-fit theory spectrum \citep{Planck2018spec}. The recovered spectrum roughly matches up until $\ell\approx 900$, after which beam effects start causing the spectrum to diverge. It also has slightly less scatter relative to the theory spectrum than the published Planck spectrum and has some minor disagreement for $\ell \lesssim 250$. This minor disagreement is not manifested when the model is used to clean Planck FFP10 simulated frequency maps \citep{Planck2018hfi}, so the source of the disagreement is not captured by these simulations. Furthermore, using frequency maps produced by the \emph{SRoll 2.2} mapmaking pipeline \citep{Delouis2019} produces no significant difference in the recovered spectrum when compared to that recovered from the Planck 2018 legacy maps, so the minor disagreement is likely not due to the instrumental effects that are better corrected for by the SRoll 2.2 pipeline. The exact cause of the minor disagreement remains unknown.

\subsection{Comparison of dust models}
\label{sec:dustmodels}

\begin{figure}
\centering
\includegraphics[width=\columnwidth]{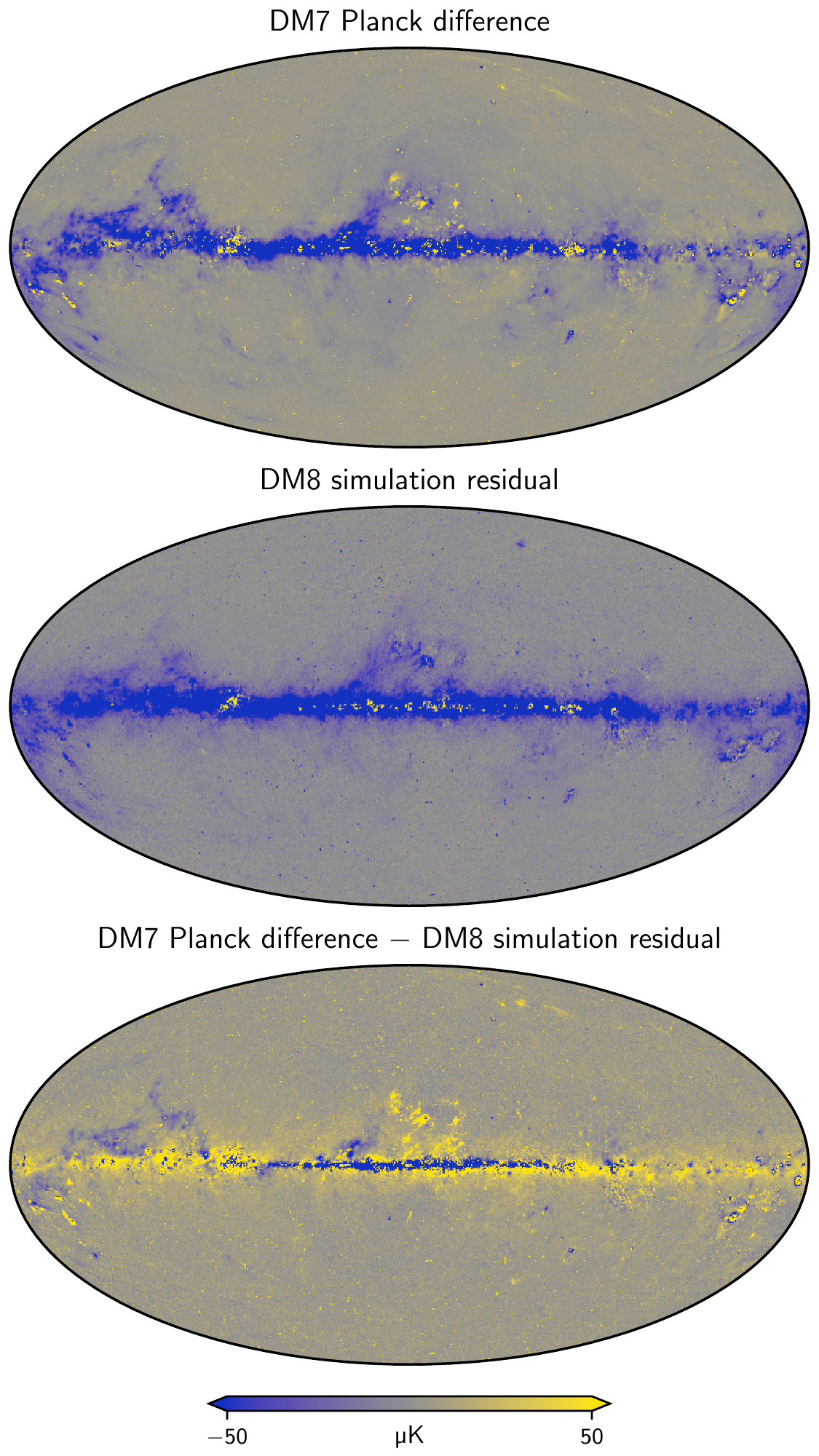}
\caption{Dust model comparison. The top panel shows the difference between the Planck Commander map and the CMB as reconstructed by the neural network model when trained on simulations that use Model 7 of \citet{Finkbeiner1999} for thermal dust. The middle panel shows the residual when the same neural network model is used to evaluate a simulation that uses Model 8 of \citet{Finkbeiner1999} for thermal dust. The bottom panel shows the difference between the top two panels. Note that it strongly resembles the difference map shown in Figure \ref{fig:planckmaps}, which was produced using the retrained model. All three panels share the same color scale; note that it saturates on portions of the Galactic plane, particularly for the top two panels.}
\label{fig:dustmodels}
\end{figure}

The neural network model described above was originally trained using Planck Sky Model simulations that modeled thermal dust using Model 7 of \citet{Finkbeiner1999}. When the neural network model was evaluated on the Planck frequency maps, this resulted in a large residual when compared to the Commander map. As qualitative visual comparisons showed a strong resemblance between this residual and maps of Galactic thermal dust foregrounds, the Planck Sky Model code was modified to use Model 8 of \citet{Finkbeiner1999} for its thermal dust simulation. The angular power spectrum resulting from evaluation of the Planck frequency maps with the neural network model did not significantly change when the dust model used for the training simulations was updated, suggesting that the minor discrepancy with the published Planck spectrum for $\ell \lesssim 250$ is not related to the dust model used in the simulations. Due to the long training process, an updated simulation was first evaluated with the existing neural network model before the model was retrained. The residual for this evaluation appeared similar to the difference between the results of the evaluation of Planck frequency maps and the Commander map, suggesting that the new simulations better matched it. The original Commander difference, the updated simulation residual, and the difference between the two are shown in Figure \ref{fig:dustmodels}.

The difference between the two residual maps strongly resembles the actual difference between the retrained model evaluated on the Planck frequency maps and the Commander map, shown in Figure \ref{fig:planckmaps}. This demonstrates the ability of the neural network model to evaluate revised simulations. Although retraining the model is computationally expensive, evaluating the trained model for new data is not. Thus, adjustments to simulations can be quickly checked against observations by comparing the residuals produced when the trained model is used to evaluate both the observations and the adjusted simulations; the closer the residuals match, the closer the adjusted simulations match the observations. Once the residuals are sufficiently close, the neural network model can be retrained on the adjusted simulations and then be used to reevaluate the observations. This process can be repeated as necessary. By only revising the simulations instead of both the simulations and the model, one can iterate on simulation differences and improvements much more quickly than would otherwise be possible. Although comparing the residuals requires an external foreground cleaning method to reference to, it has the advantage of providing hints as to what needs to be changed in the simulations. The neural network model alone also provides a metric by which to evaluate the fidelity of the simulations, its mean predicted error. This went from \SI{11.2}{\micro\kelvin} to \SI{7.4}{\micro\kelvin} without any sky regions masked, again showing the improvement of using the revised thermal dust model.

\section{Conclusions}
\label{sec:conclusion}

Using a Bayesian spherical convolutional neural network, a machine-learning foreground cleaning technique was developed for CMB observations. The model was trained on full-sky temperature simulations that incorporated knowledge of the millimeter and submillimeter sky that was available at the time of the launch of the Planck satellite. After being shown to work on simulations, the trained model was applied to Planck observations. The neural network code, trained weights, and results have been made available \citep{Petroff2020}.

As the current neural network model is trained on simulations, the effectiveness of the foreground cleaning is limited by the accuracy of the simulations. Thus, the effectiveness of the cleaning should improve if the training simulations are improved. While the Planck Sky Model is a template-based phenomenological method, the use of more physically motivated simulations, e.g., along the lines of what \citet{Puglisi2017} developed for diffuse Galactic CO emission, could potentially be an improvement, although the development of such foreground simulations is outside the scope of the present work. The demonstrated ability of the current neural network model to evaluate how well different simulations match observations or different simulations could potentially aid in the development of these improved simulations as it can quickly provide a metric for evaluating how well the simulation matches and a residual map that can hint at what parts of the simulation still needs to be improved.

In this work, only temperature maps were considered, so one avenue of future work is to extend the presented technique to polarization maps. In early experiments, the presented technique was found to perform poorly on Planck-like polarization maps. However, as the technique did work when the noise levels on the simulated polarization maps were reduced, the deficiency seems to be due to the inability to handle the higher noise levels in the Planck polarization maps, which caused the loss function to be best minimized by smoothing the maps and eliminating small angular scale features.  Potential solutions to this issue include incorporating the angular power spectrum difference into the loss function or using generative adversarial networks (GANs) \citep{Isola2017}.

\vskip 5.8mm plus 1mm minus 1mm
\vskip1sp
\section*{Acknowledgments}
\vskip4pt

This research was supported in part by NASA grant numbers 80NSSC19K0526 and 80NSSC20K0445. We acknowledge the support of the National Science Foundation Division of Astronomical Sciences under grant number 1636634. We thank an anonymous referee for suggestions that helped improve this manuscript. This work is in part based on observations obtained with Planck, an ESA science mission with instruments and contributions directly funded by ESA Member States, NASA, and Canada. We acknowledge the use of the Legacy Archive for Microwave Background Data Analysis (LAMBDA), part of the High Energy Astrophysics Science Archive Center (HEASARC); HEASARC/LAMBDA is a service of the Astrophysics Science Division at the NASA Goddard Space Flight Center. This research has made use of NASA's Astrophysics Data System.

\software{Planck Sky Model v1.7.8 \citep{Delabrouille2013}, CLASS v1.5 \citep{Blas2011}, PolSpice v03-05-02 \citep{Chon2004}, Healpy v1.12.10 \citep{Zonca2019, Gorski2005}, \textsc{Healpix} v3.50 \citep{Gorski2005}, TensorFlow v2.1.0 \citep{Abadi2016}, TensorFlow Large Model Support \citep{Le2019}, NumPy v1.17.2 \citep{vanderWalt2011}, Matplotlib v3.1.1 \citep{Hunter2007}}

\bibliography{paper.bib}

\end{document}